\documentstyle[epsf,german]{article}
\newcommand{\be}{\begin{eqnarray}}
\newcommand{\ee}{\end{eqnarray}}

\newcommand{\bi}{\bigskip}
\newcommand{\no}{\noindent}
\newcommand{\hk}{\hspace{0.1cm}}

\newcommand{\rk}{\right)}
\newcommand{\lk}{\left(}

\newcommand{\sli}{\sum\limits}

\newcommand{\il}{\int\limits}

\begin{document}
\centerline{\bf Magnetic Monopoles, Center Vortices, Confinement 
and Topology of Gauge Fields\footnote{invited talk given by H. Reinhardt on the
Int. Workshop ``Hadrons 1999'', Coimbra, 10.-15. Sept. 1999}
}  
\centerline{H. Reinhardt, M. Engelhardt, K. Langfeld, M. Quandt, A. Sch\"afke}
\centerline{Institut f\"ur Theoretische Physik} 
\centerline{Universit\"at T\"ubingen,
 Auf der Morgenstelle 14}
\centerline{D-72076 T\"ubingen, 
Germany} 

\bi

\no
\begin{abstract}
The vortex picture of confinement is studied. The deconfinement phase transition
is explained as a transition from a phase in which vortices percolate to a phase
of small vortices. Lattice results are presented in support of this scenario.
Furthermore
the topological properties of magnetic monopoles and center vortices arising,
respectively, in Abelian and center gauges are studied in continuum
Yang-Mills-theory.
For this purpose the continuum analog of the maximum center gauge is
constructed.
\end{abstract}
\section{Introduction}

Recent lattice calculations have given strong evidence for two confinement
scenarios: 1. the dual Meissner effect \cite{1}, which is based on a condensate
 of
magnetic monopoles in the QCD vacuum and 2. the center vortex picture \cite{2},
 where the
vacuum consists of a condensate of magnetic flux tubes which are closed due 
to the
Bianchi identity. There are also lattice calculations which indicate that the
spontaneous breaking of chiral symmetry, which can be related to the topology of
gauge fields, is caused by these objects, i.e. by either magnetic monopoles
\cite{3} or
center vortices \cite{4}. 
In this talk we would like to discuss the confinement and topological properties
of magnetic monopoles and center vortices. We will first discuss the two
confinement scenarios based on magnetic monopoles and vortices, respectively,
and subsequently investigate the topological properties of these objects. In
particular, we will study the nature of the deconfinement phase transition in
the center vortex picture. We
will also show that in Polyakov
gauge the magnetic monopoles completely account for the non-trivial topology of
the gauge fields. Subsequently, we will extend the notion of center vortices 
to the
continuum. We will present the continuum analog of the maximum center gauge
fixing and the Pontryagin index of center vortices.
\bi

\no
\section{Confinement}
\bi

\no
The magnetic monopoles arise in Yang-Mills-Theories in the so called Abelian
gauges \cite{5}.
Recent lattice calculations have shown that below a critical temperature $T_C$
these monopoles are condensed \cite{R1} and give rise to the dual Mei"sner
effect. In particular in the so called maximal Abelian gauge where all links are
made as diagonal as possible, one observes Abelian and monopole dominance in
the string tension \cite{1}. However, very recent lattice calculations \cite{R2}
also show that the Yang-Mills ground state does not look like a Coulombic
monopole gas but rather indicate a collimation of magnetic flux, which is
consistent with the center vortex picture of confinement, proposed in refs.
\cite{X3}, \cite{X4}, \cite{X6}, \cite{X5}.

Center vortices represent closed magnetic flux lines in three space dimensions,
describing closed two-dimensional world-sheets in four space-time dimensions.
The magnetic flux represented by the vortices is furthermore quantized such that
a Wilson loop linking vortex flux takes a value corresponding to a nontrivial
center element of the gauge group. In the case of $SU (2)$ colour 
the only such element is (-1). For $N$ colours, there are $N - 1$
different possible vortex fluxes corresponding to the $N - 1$ nontrivial center
elements of $SU (N)$. Center vortices can be regarded as a
fraction of a Dirac string: $N$ superimposed 
center vortices form an unobservable Dirac
string.

Consider an ensemble of center vortex configurations in which the vortices are
distributed randomly, specifically such that intersection points of vortices
with a given two-dimensional plane in space-time are found at random,
uncorrelated locations. In such an ensemble, confinement results in a very
simple manner. Let the universe be a cube of length $L$, and consider a
two-dimensional slice of this universe of area $L^2$, with a Wilson loop
embedded into it, circumscribing an area $A$. On this plane, distribute $N$
vortex intersection points at random, cf. Fig. 1 (left). According to the
specification above, each of these points contributes a factor (- 1) to the
value of the Wilson loop if it falls within the area $A$ spanned by the loop;
the probability for this to occur for any given point is $A/L^2$.

\begin{figure}[t]
\centerline{
\vspace{2cm} 
\hspace{2cm}
\epsfxsize=3cm
\epsffile{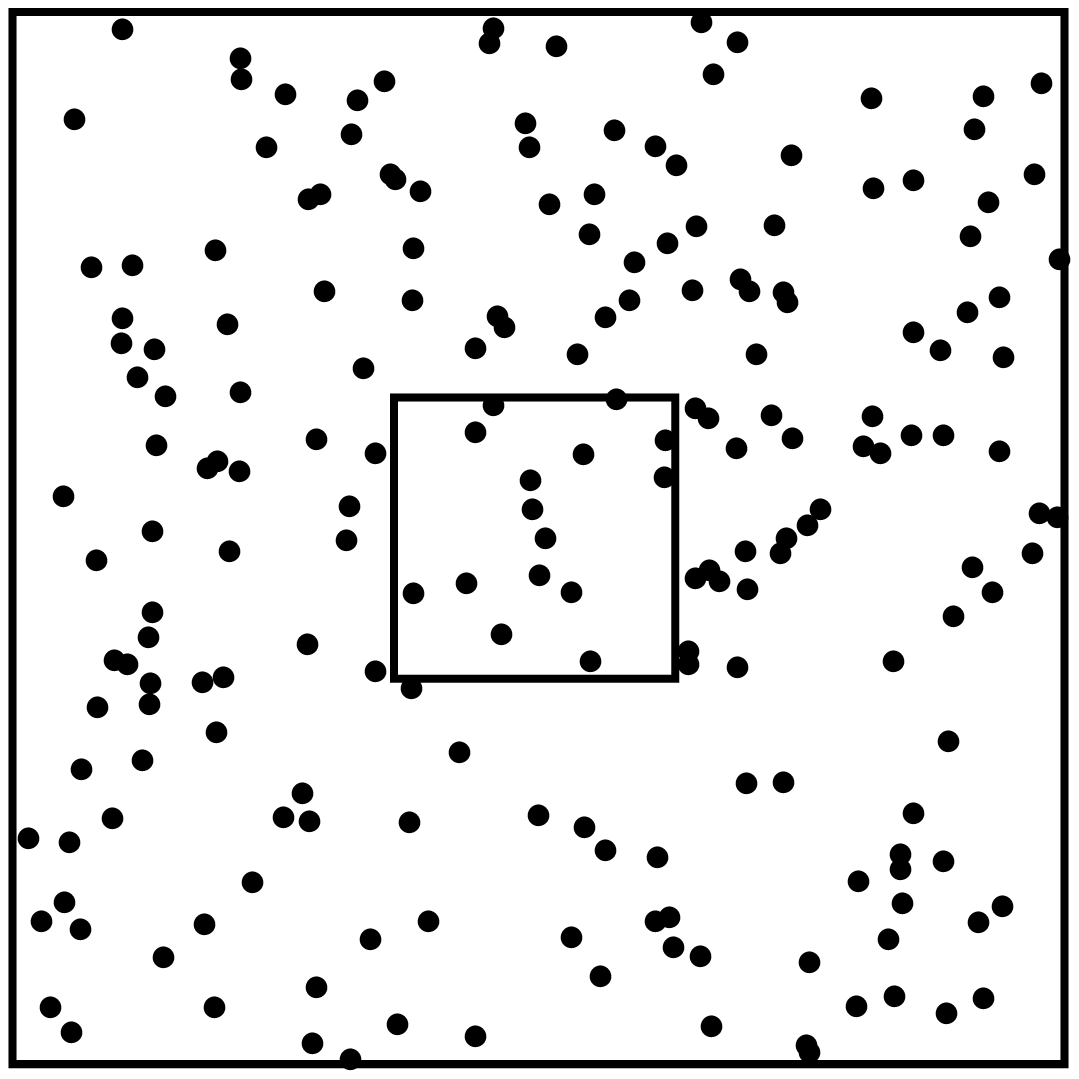}
\hspace{2cm}
\epsfxsize=3cm
\epsffile{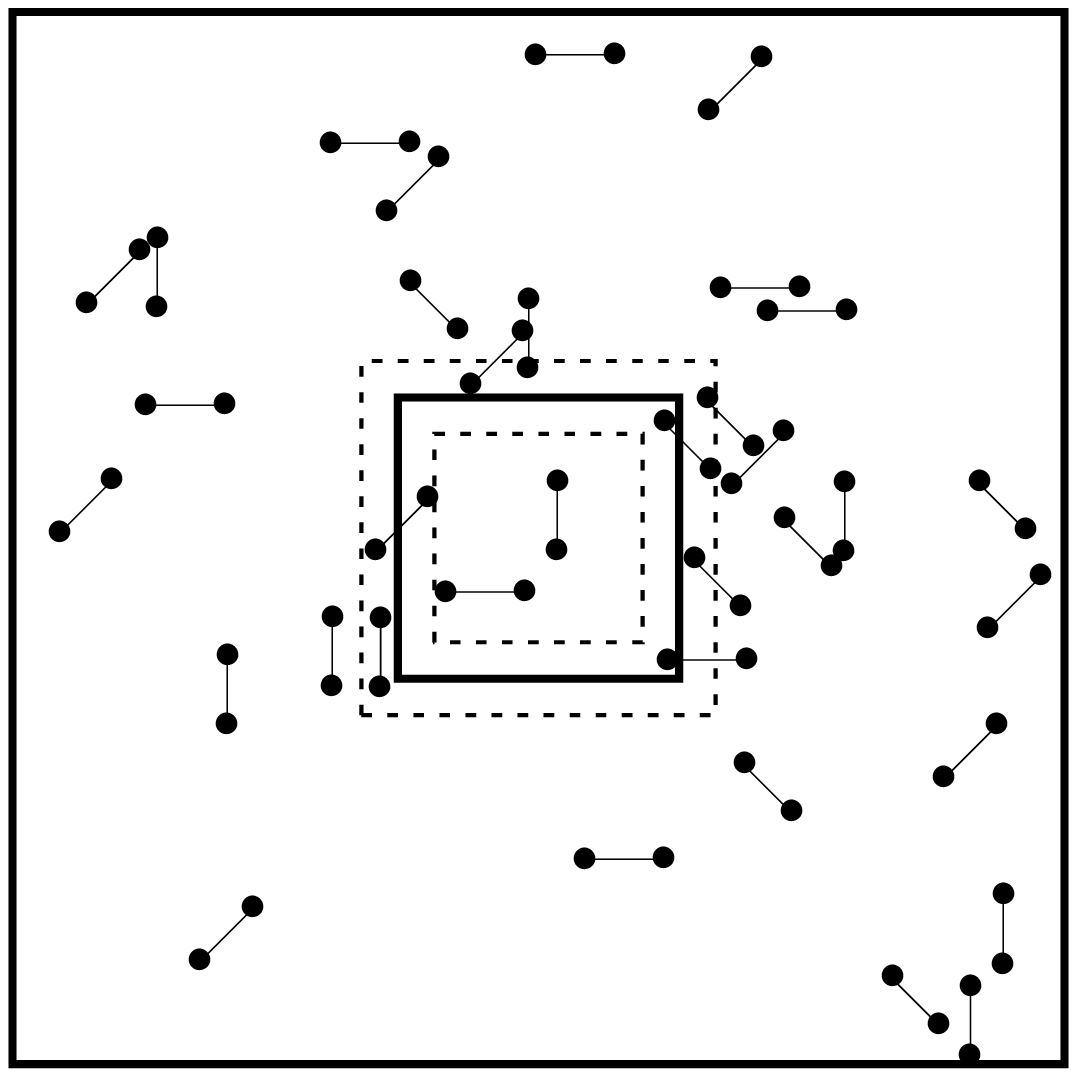}
}
\caption{Two-dimensional plane of a random vortex gas with a planar Wilson loop.
The dots represent the intersection points of the vortices with the plane
considered. Left figure: confining vortex ensemble of uncorrelated intersection
points. Right figure: deconfining vortex ensemble in which the intersection
points occur pairwise within a distance $d$ away from each other. (The
intersection points of a pair are connected by a line.) } 
\label{fig:e1} 
\end{figure}

The expectation value of the Wilson loop is readily evaluated in this simple
model. The probability that $n$ of the $N$ vortex intersection points fall
within the area $A$ is binomial, and, since the Wilson loop takes the value $(-
1)^n$ in the presence of $n$ intersection points within the area $A$, its
expectaton value is
\be
\label{Y1}
<W> & = & \sli^N_{n = 0} (- 1)^n \lk
\begin{array}{c}
N \\
n 
\end{array} \rk
\lk \frac{A}{L^2} \rk^n \lk 1 - \frac{A}{L^2} \rk^{N - n} \nonumber\\
& = & \lk 1 - \frac{2 \rho
A}{N} \rk^N \stackrel{N \to \infty}{\longrightarrow} exp (- 2 \rho A) \hk ,
\ee
where in the last step, the size of the universe $L$ has been sent to infinity
while leaving the planar density $\rho  = N /L^2$ of vortex intersection points
constant. Thus, one obtains an area law for the Wilson loop, with the string
tension $\sigma_{r v m} = 2 \rho$.

In fact, in lattice calculations the vortex area density $\rho$ has been shown
to obey the proper scaling behaviour \cite{X1} 
as dictated by the renormalization group
and thus represents a physical observable. Using a string tension of $\sigma
\sim (440 MeV)^2$ as input one finds $\rho \approx 3.4 
F^{- 2}$ corresponding to a
string tension $\sigma_{r v m} = (521 MeV)^2$ in the random model above 
which overestimates the input value.
This overabundance of string tension can be easily understood by noticing that there
are both dynamical \cite{X8} and kinematical correlations between the vortex
intersection points, which have been discarded in the random vortex model
considered above, which assumes that all intersection points are completely
random, i.e. uncorrelated. This is, however, not true since the vortices are
closed loops in $D = 3$ or closed surfaces in $D = 4$. Therefore the
intersection points in the plane of the Wilson loops come in pairs. But a pair
of intersection points does not (non-trivially) contribute to the Wilson loop.
Only for large vortices exceeding the size of the Wilson loop the intersection
points inside the Wilson loop are uncorrelated and can contribute $(- 1)$. On
the other hand all vortices contribute to the area vortex density $\rho$
measured on the lattice. This effect leads to a lower value of the string
tension than the value $\sigma_{r v m} = 2 \rho$ resulting from 
the random vortex model.
\bi

\no
\section{Deconfinement}
\bi

\no
The above presented
vortex picture of confinement naturally explains also the deconfinement
transition as a transition from a phase of large vortices percolating throughout
space-time to a phase of small vortices in a sense to be specified more
precisely below. Indeed, assume that all vortices have a
maximal size $d$. Then only the intersection points in a strip of width $d$
along the perimeter of the Wilson loop can randomly
contribute $(- 1)$, (while other intersection points come in pairs and hence do
not contribute). The expectation value of the Wilson loop is then still given by
eq. (\ref{Y1}), however, with the full area $A$ of the Wilson loop replaced by the area
of the strip of width $d$ along the perimeter $P$ of the Wilson loop, $d
\cdot P$, resulting in a perimeter law
\be
<W> = exp (- 2 \rho d P)
\ee
implying deconfinement. This picture of the deconfinement phase transition
arising in the random vortex model as a transition from a phase of percolating
vortices to a phase of small vortices is supported by the lattice calculations.
Fig. 2 shows the vortex matter distribution as function of the vortex cluster
extension at various temperatures for a 3-dimensional slice resulting from the
4-dim. lattice by omitting one spatial direction\cite{20}, 
see also ref. \cite{X8}. 
Far below the critical temperature $T_C$ of the deconfinement phase
transition a dominant portion of the 
vortex matter is contained in a big cluster
extending over the whole lattice universe. As the temperature rises smaller
clusters are more and more formed and well above the deconfinement phase
transition 
large vortices 
have ceased to exist, the connectivity of the clusters is lost and 
all vortex matter being contained in small clusters.

\begin{figure}[t]
\centerline{ 
\epsfxsize=4cm
\epsffile{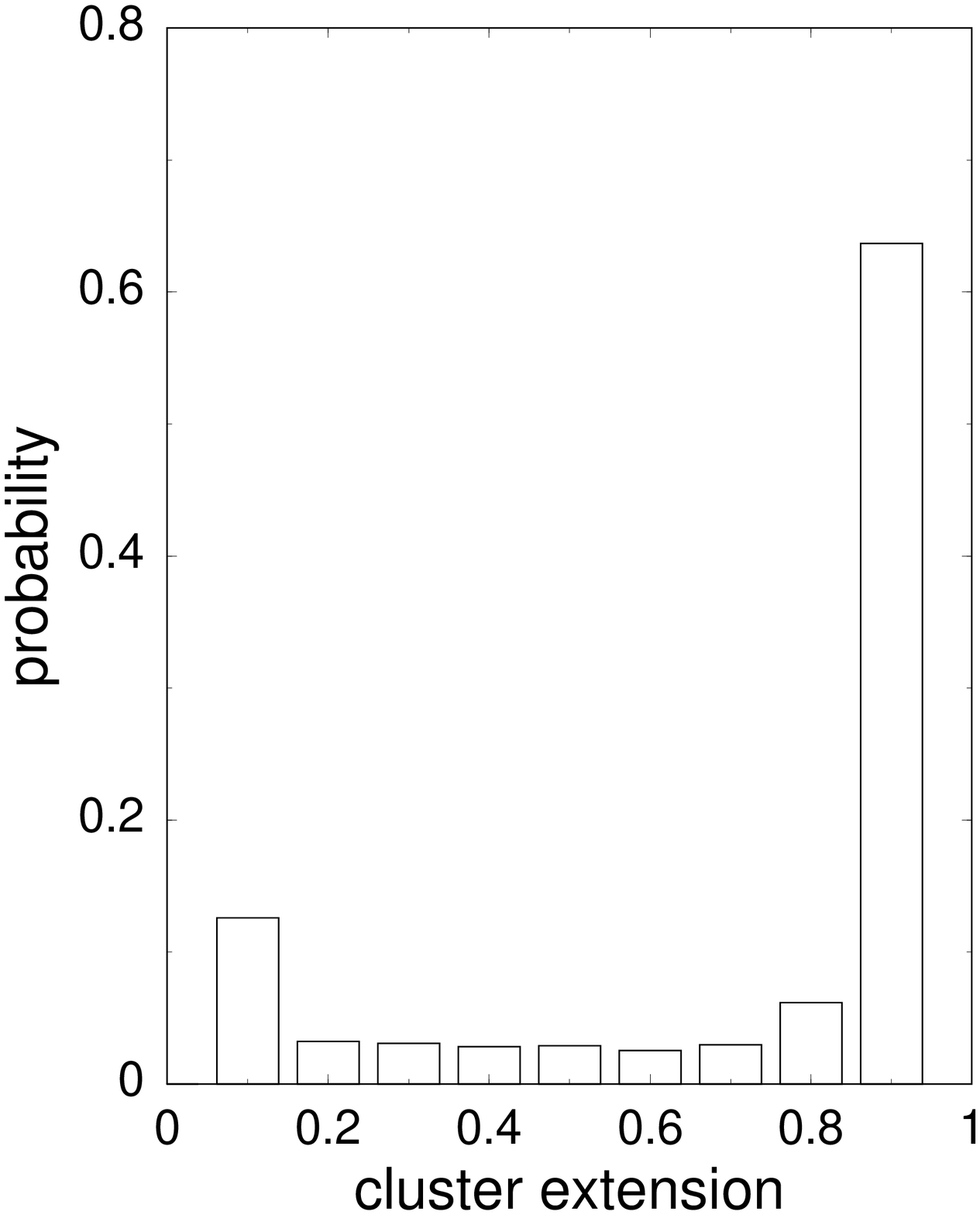}
\hspace{1cm}
\epsfxsize=4cm
\epsffile{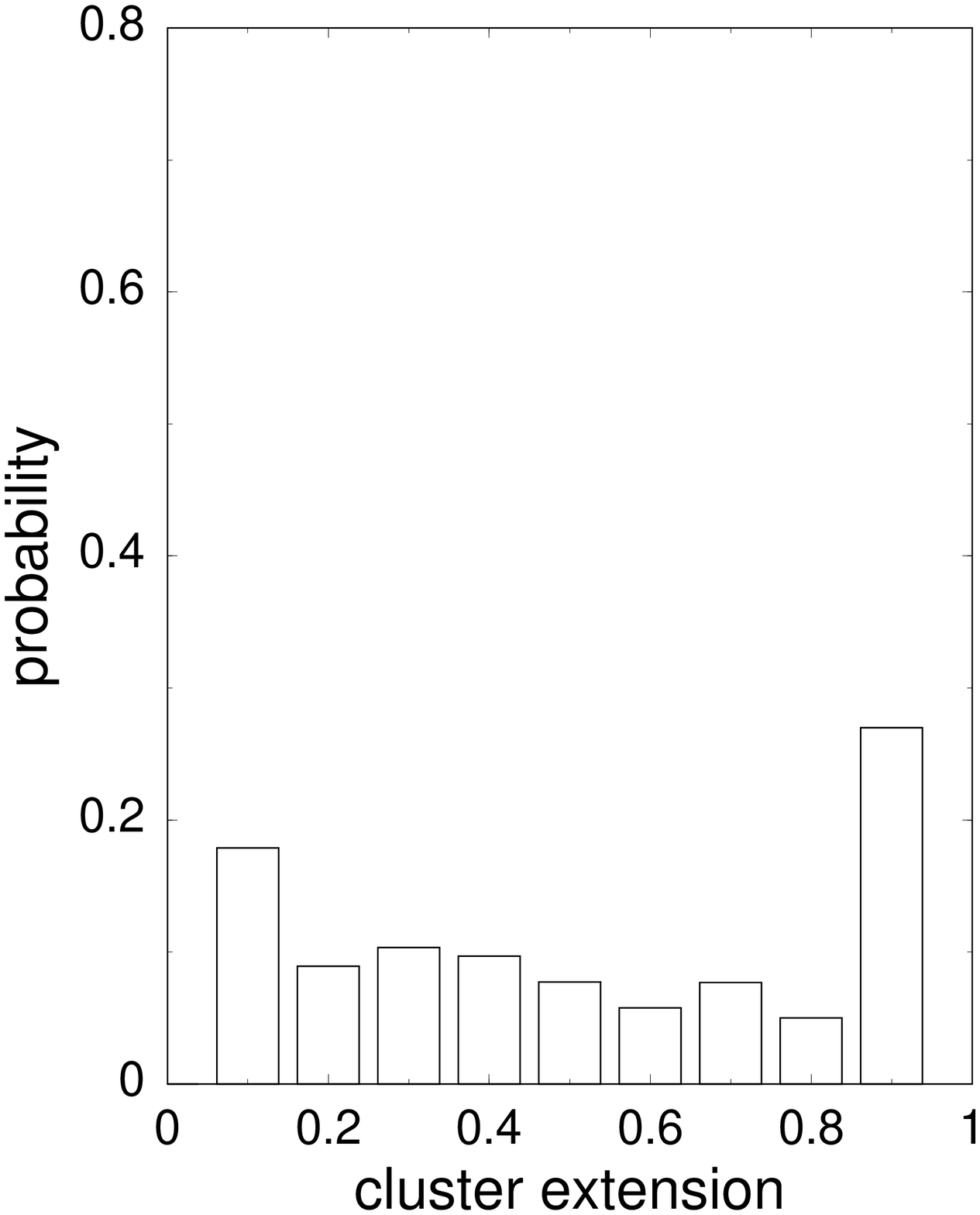}
}
\centerline{ 
\epsfxsize=4cm
\epsffile{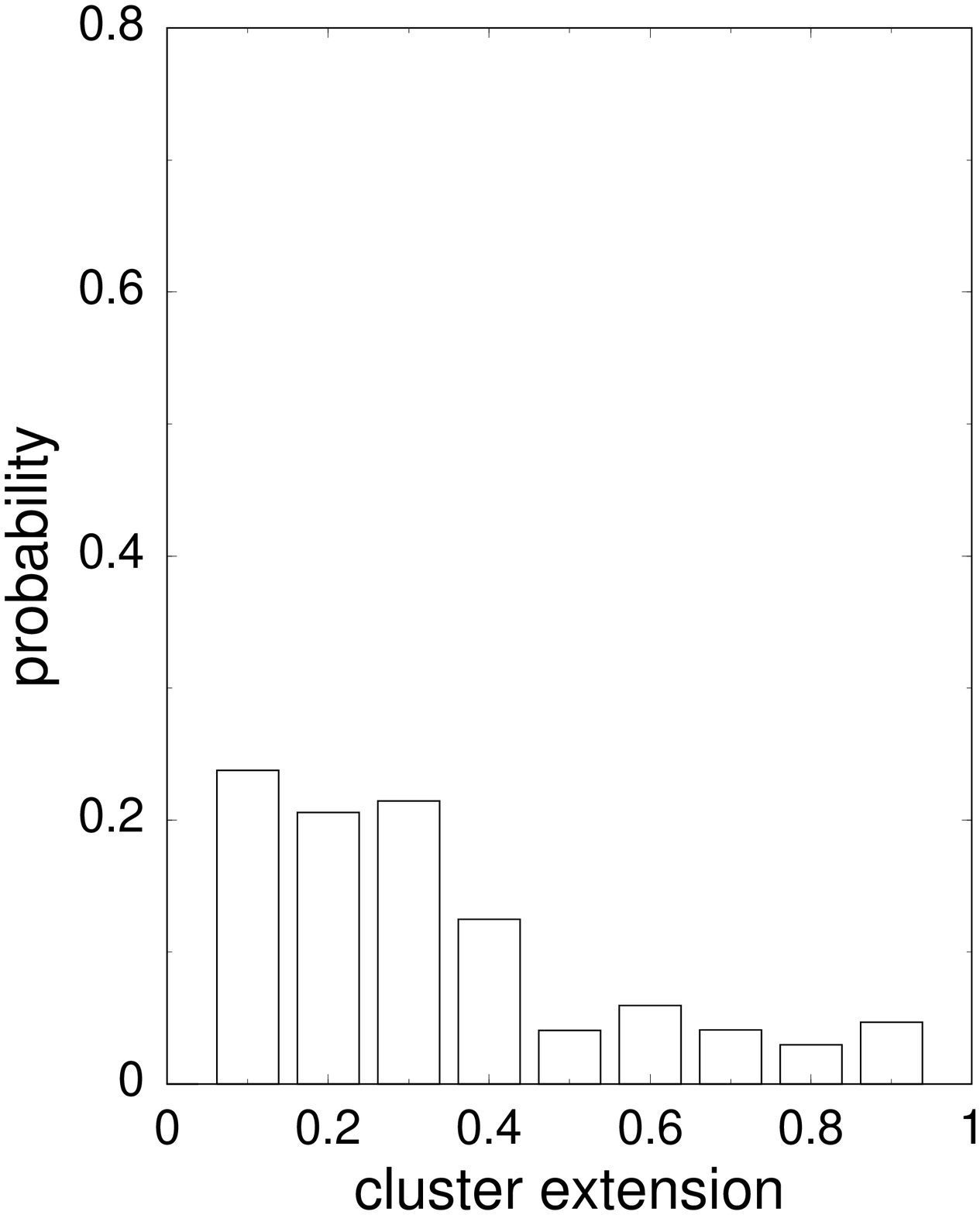}
\hspace{1cm}
\epsfxsize=4cm
\epsffile{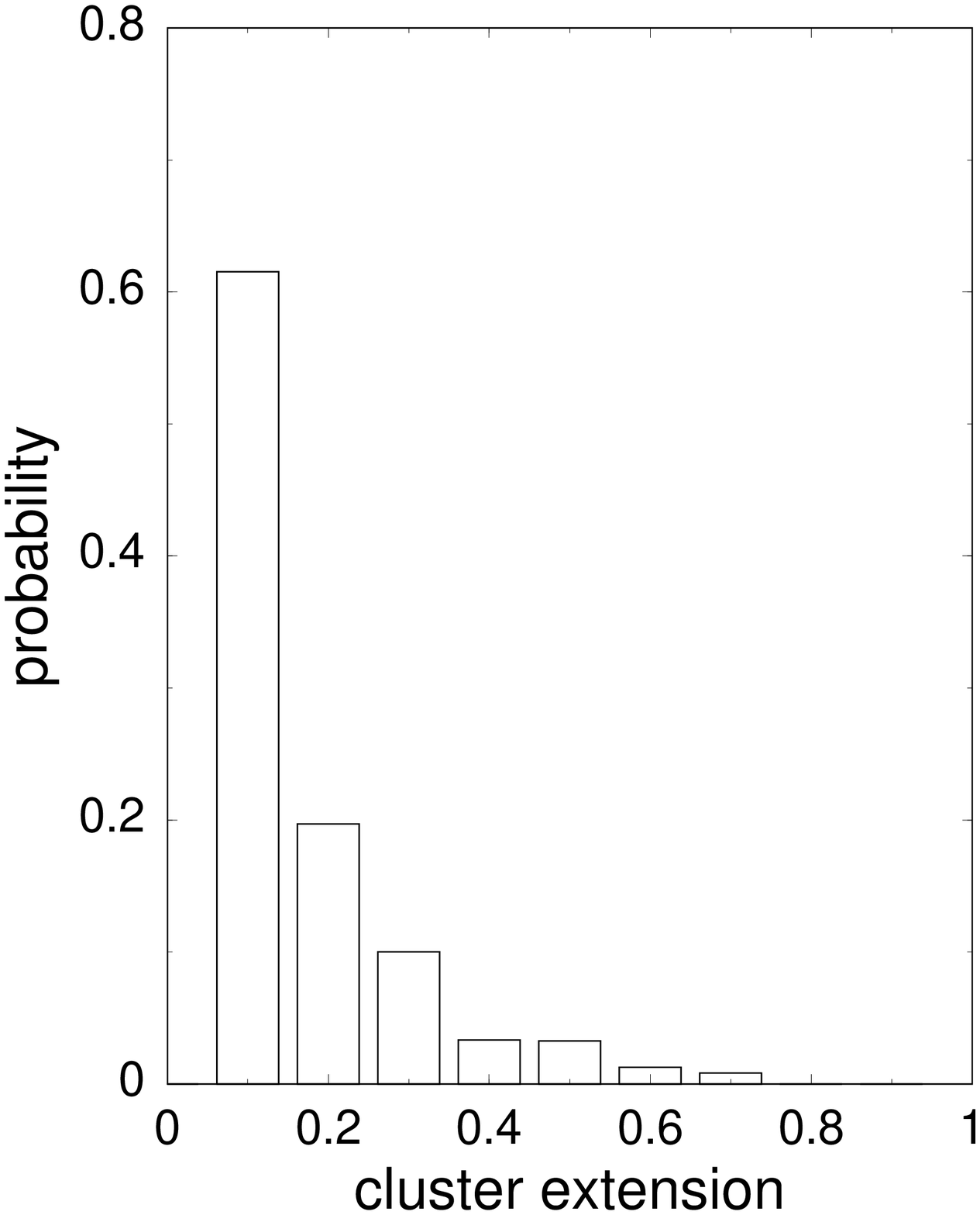}
}
\caption{ Vortex material distributions. } 
\label{fig:e2} 
\end{figure}

If one analyzes the small vortex clusters dominating the deconfined phase in
more detail, one finds that a large part of these vortices wind in the
(Euclidean) temporal direction, i.e. the space-time direction whose extension is
identified with the inverse temperature. Therefore, one finds that the typical
configurations in the two phases can be characterized as displayed in Fig. 3 in
a three-dimensional slice of space-time, where one space direction has been left
away. Note that Fig. 3 also furnishes an explanation of the spatial string
tension in  the deconfined phase. A spatial Wilson loop embedded into Fig. 3
(right) can exhibit an area law since intersection points of winding vortices
with the minimal area spanned by the loop can occur in an uncorrelated fashion
despite those vortices having small extension. Note also the dual nature of this
(magnetic) picture as compared with electric flux models \cite{X12}. In such
models, electric flux percolates in the {\it deconfined} phase, while it does
not percolate in the confining phase.

\begin{figure}[t]
\centerline{ 
\epsfxsize=4cm
\epsffile{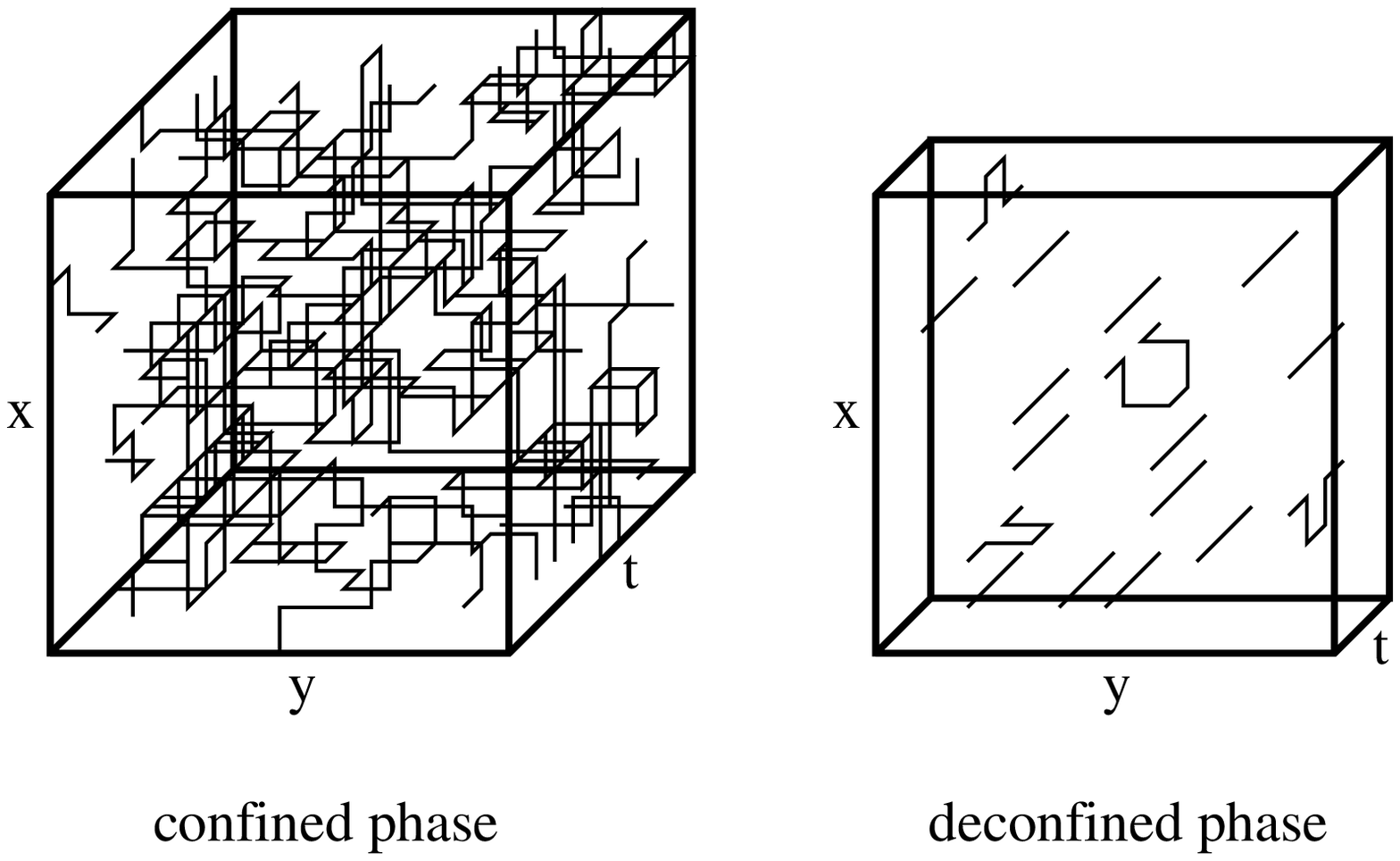}
}
\caption{ Typical vortex configurations in the confining (left) and the
deconfined phase (right). } 
\label{fig:e3} 
\end{figure}
\bi

\no
\section{Magnetic monopoles and topology}
\bi

\no
Spontaneous breaking of chiral symmetry can be triggered by topologically
non-trivial gauge fields, which give rise to zero modes of the quarks \cite{21}.
Magnetic monopoles and percolated vortices are long range fields and should
hence be relevant for the global topological properties.
\bi

\no
Topological properties of gauge configurations as measured by the Pontryagin
index
\be
\nu = \frac{- 1}{16 \pi^2} \int T r F_{\mu \nu} \tilde{F}_{\mu \nu} = \frac{1}{4
\pi^2} \int d^3 x \vec{E} (x) \vec{B} (x)
\ee
are preferably studied in the continuum theory.
For the study of the topological properties of magnetic 
monopoles in the continuum theory the
Polyakov gauge is particularly convenient. In this gauge one diagonalizes 
the Polyakov loop  
\begin{eqnarray}
\label{1}
\Omega(\vec{x}) = Pe^{- \int^{T}_{0} dx_{0} A_{0}(x_{0}, \vec{x})} = V^{\dagger}
\omega V
\end{eqnarray}
which fixes $V \in SU(2) / U(1)$ i.e. the coset part of the gauge group, which
we assume, for simplicity, to be $SU (2)$.
Magnetic monopoles arise as defects of the gauge fixing, which occur when at 
isolated points in space $\vec{x}_{i}$ the Polyakov loop becomes 
a  center element
\begin{eqnarray}
\label{2}
\Omega(\vec{x}_{i}) = (-1)^{n_i} , \; \; \;  n_{i}\, :\, \hbox{integer} 
\end{eqnarray}
The field $A^{V} =  V A V^{\dagger} + V \partial V^{\dagger}$ 
develops then  magnetic 
monopoles located at these points. 
These monopoles have topologically quantized magnetic charge \cite{6} given by 
the winding number
\begin{eqnarray}  
\label{3}
m[V] \in \Pi_{2} (SU(2)/U(1))     
\end{eqnarray}
of the mapping $V(\vec{x})$ from a sphere $S_2$ around the magnetic 
monopole into the coset $SU(2)/U(1)$ 
of the gauge group. 

In the Polyakov
gauge the Pontryagin index can  be exactly expressed in terms of magnetic
charges \cite{6}, \cite{22}, \cite{X9}. 
If we assume a compact space-time mani\-fold and that there are 
only point-like defects of gauge fixing,
i.e magnetic monopoles are the only magnetically charged objects arising after
gauge fixing, the Pontryagin index is given by \cite{6}
\begin{eqnarray}
\label{7a}
\nu = \Sigma_{i} n_i m_i
\end{eqnarray}
The summation runs here over all magnetic monopoles with $m_i$ being the
magnetic charge of the monopole defined by equation (\ref{3}) and the integer 
$n_i$ is defined by the value of the Polyakov-loop at the monopole position
(\ref{2}).
This relation shows that the magnetic monopoles completely account for the 
non-trivial topology of gauge fields, at least in the Polyakov gauge. 
Unfortunately,
in other Abelian gauges like maximum Abelian gauge, such a simple relation
between Pontryagin index and magnetic charges is not yet known and perhaps
does not exist \cite{X9}. However, in
the maximum Abelian gauge correlations between instantons and monopoles have
 been
found, in both analytical and lattice studies \cite{3}. 

\section{Center vortices in the continuum}

On the lattice center vortices are detected by going to the maximum center gauge
and subsequently projecting the links onto center elements \cite{2}. 
In the maximum center gauge 
\be \sum_{x,{\mu}} (T r U_{\mu}(x))^{2} 
\to max \hk  , 
\ee 
which is obviously insensitive to center gauge transformations, one exploits
the gauge freedom to 
rotate a link variable as close as possible to a center element. 
Once the maximum center gauge has been implemented, center
projection implies to replace all links by their closest center element. 
One obtains then a
$Z(2)$ lattice which contains $D - 1$ dimensional hypersurfaces $\Sigma$  on 
which all links take a non-trivial center element, that is $U = -1$ in the 
case of
$SU(2)$. The $D - 2$ dimensional boundaries $\partial\Sigma$ of the 
hypersurfaces
$\Sigma$ represent the center vortices, which, when non-trivially linked to a
Wilson loop, yield a center element for the latter. The notion of center 
vortices can be extended to the continuum theory by putting a 
given smooth gauge field 
$A_{\mu}(x)$
on a lattice in the standard fashion by introducing the link variables
$U_{\mu}(x) = exp(-aA_{\mu}(x))$.

A careful analysis shows that the continuum analogies of the center 
vortices are defined  by the gauge potential \cite{7},
\begin{eqnarray}
\label{4}
{\cal A}_{\mu}(x,\Sigma) = E \int_{\Sigma} d^{D-1} \tilde{\sigma}_{\mu} \delta^{D}(x -
\bar{x}(\sigma))      
\end{eqnarray}
where $d^{D-1} \tilde{\sigma}_{\mu}$ is the dual of the $D-1$
dimensional volume element. Furthermore, the quantity $E = E_{a}H_{a}$ with
$H_{a}$ being the generators of the Cartan algebra represents 
(up to a factor of
$2\pi$) the so called co-weights which satisfy $exp (-E) = Z \in Z(N)$.
Due to this fact the Wilson-loop calculated from the gauge potential (\ref{4})
becomes,
\begin{eqnarray}
\label{5}
W[{\cal A}](C) = \exp (- \oint_{C} {\cal A}) = Z^{I(C,\Sigma)}
\end{eqnarray}
where $I(C,\Sigma)$  is the intersection number between the Wilson-loop $C$ and
the hypersurface $\Sigma$. The representation, (\ref{4}), 
is referred to as ideal
center vortex. One should emphasize that the hypersurface $\Sigma$ can be
arbitrarily deformed by a center gauge transformation keeping, however, 
its boundary
$\partial\Sigma$, i.e. the position of the center vortex, fixed. Thus for fixed 
$\partial\Sigma$  the dependence of the gauge potential (\ref{4}) on the 
hypersurface itself is a gauge artifact. 

The dependence on the hypersurface $\Sigma$ can be removed by performing the
gauge transformation 
\begin{eqnarray}
\label{6}
\varphi(x,\Sigma) = exp (- E \Omega (x,\Sigma))
\end{eqnarray}
where $\Omega(x,\Sigma)$ is the solid angle subtended by the hypersurface
$\Sigma$ as seen from the point $x$. One finds then
\begin{eqnarray}
\label{7}
{\cal A}_{\mu}(x, \Sigma) = \varphi (x,\Sigma) \partial_{\mu}
\varphi^{\dagger}(x,\Sigma) + a_{\mu} (x,\partial\Sigma)
\end{eqnarray}
where
\begin{eqnarray}
\label{8}
a_{\mu}(x, \partial\Sigma) = E \int_{\partial\Sigma} d^{D-2}
\tilde{\sigma}_{\mu\nu} \partial_{\nu} D (x - \bar{x}(\sigma))    
\end{eqnarray}
depends only on the vortex position $\partial\Sigma$ and is referred to 
as ''thin
vortex''. Here $D(x -\bar{x}(\sigma))$ represents the Green function 
of the $D$
dimensional Laplacian. In fact, one can show \cite{7} that the thin vortex 
represents the
transversal part of the ideal vortex $a_{\mu}(x, \partial\Sigma) = P_{\mu\nu}
{\cal A}_{\nu} (x, \Sigma)$ where $P_{\mu\nu} = \delta_{\mu\nu} - \frac
{\partial_{\mu}\partial_{\nu}}{\partial^{2}}$ is the usual transversal
projector.
A careful and lengthy analysis \cite{7} yields that 
the continuum
analog of the maximum center gauge fixing is defined by 
\begin{eqnarray}
\label{9}
\min\limits_{\partial\Sigma} \min\limits_{g} \int \hbox{Tr}  (A^{g} - 
a (\partial\Sigma))^{2}
\end{eqnarray}
where the minimization is performed with respect to all (continuum) gauge
transformations $g \in SU(2)/Z(2)$ (which represent per se coset gauge
transformations) and with respect to all vortex surfaces $\partial\Sigma$. For
fixed thin center vortex field configuration $a (\partial\Sigma)$ 
minimization with
respect to the continuum gauge transformation g yields the background gauge
condition 
\begin{eqnarray}
\label{10}
\left[\partial_{\mu} + a_{\mu} (\partial\Sigma), A_{\mu} \right]
 = 0
\end{eqnarray}
where the thin vortex field $a_{\mu} (x, \partial\Sigma)$ figures as background
gauge field. One should emphasize, however, that the background field has to be
dynamically determined for each given gauge field $A_{\mu} (x)$ and thus depends
on the latter.
Obviously in the absence of a vortex structure in the considered gauge field 
$A_{\mu} (x)$ the background gauge condition reduces to the Lorentz gauge
$\partial_{\mu}  A_{\mu} = 0$.

\section{Topology of Center vortices}

Once  the center vortex configurations 
in the continuum are at our disposal, it 
is straightforward to calculate their Pontryagin index. 
In the continuum formulation where center vortices live in the
Abelian subgroup by construction the direction of the magnetic flux of 
the vortices is fully kept.
The explicit calculation \cite{7} 
shows that the Pontryagin index $\nu$ of the center vortices
is given by 
\be
\nu = \frac{1}{4}
I(\partial\Sigma,\partial\Sigma)
\ee
where 
$I(\partial\Sigma,\partial\Sigma)$
represents the self-intersection number of the closed vortex sheet
$\partial\Sigma$ defined by
\be
I (\partial \Sigma_1, \partial \Sigma_2) = \frac{1}{2} \il_{\partial \Sigma_1} d
\sigma_{\mu \nu} \il_{\partial \Sigma_2} d \tilde{\sigma}_{\mu \nu} \delta^4 \lk
\bar{x} (\sigma) - \bar{x} (\sigma') \rk \hk. 
\ee
A careful analysis shows that for closed oriented surfaces
the self intersection number vanishes. In order to have a non-trivial Pontryagin
index the vortex surfaces have to be not globally oriented, i.e., they have
to consist of oriented pieces. One can further show that at the border between
oriented vortex patches magnetic monopole currents flow. It is these monopole
currents which make the vortex sheet non-oriented since they change the
orientation of the magnetic flux. Thus we obtain that even for the center
vortices the non-trivial topology is generated by magnetic monopole currents
flowing on the vortex sheets. This is consistent with our finding in the
Polyakov gauge (see eq. (\ref{7a})) 
where the Pontryagin index was exclusively expressed in terms of
magnetic monopoles \cite{6}. In fact, for a compact space-time manifold one can
show that under certain mild assumptions 
the Pontryagin index can be expressed as
\be
\nu = - \frac{1}{4} L (\partial \Sigma, C) \hk ,
\ee
where $L (\partial \Sigma, C)$ is the linking number between the vortex
$\partial \Sigma$ and the monopole loop $C$ on the vortex.

By implementing the maximum center gauge condition in the continuum one 
can derive, in an approximate
fashion, an effective vortex theory \cite{7}, where the vortex action can be 
calculated in
a gradient expansion. The leading order gives the Nambu-Goto action while in
higher orders curvature terms appear. A model based on such an effective vortex
action, in fact, reproduces the gross features of the center vortex picture 
found in numerical Yang-Mills lattice simulations.

\section*{ Acknowledgment:}

This work was supported in part by the DFG-Re 856/4-1 and DFG-En 415/1-1.

	
\end{document}